\documentclass[12pt,preprint]{aastex}
\shorttitle{What Does Dimming Tell Us?}
\shortauthors{Qiu}

\begin{document}
\title{Gradual Solar Coronal Dimming and Evolution of Coronal Mass Ejection in the Early Phase}

\author{Jiong Qiu $^1$, Jianxia Cheng $^2$}
\affil{$^1$Department of Physics, Montana State University,
Bozeman, MT 59717, USA} \email{qiu@physics.montana.edu}
\affil{$^2$CAS Key Laboratory of Planetary Sciences, Shanghai
Astronomical Observatory, Chinese Academy of Sciences,
Shanghai 200030, China}
\begin{abstract}
We report observations of a two-stage coronal dimming in an eruptive event
of a two-ribbon flare and a fast Coronal Mass Ejection (CME).
Weak gradual dimming persists for more than half an hour before
the onset of the two-ribbon flare and the fast rise of the CME. It
is followed by abrupt rapid dimming. The two-stage dimming 
occurs in a pair of conjugate dimming regions adjacent to the two 
flare ribbons, and the flare onset marks the transition between 
the two stages of dimming. At the onset of the two-ribbon flare, 
transient brightenings are also observed inside the dimming regions, 
before rapid dimming occurs at the same places. These observations 
suggest that the CME structure, most probably anchored at the twin 
dimming regions, undergoes a slow rise before the flare onset, 
and its kinematic evolution has significantly changed at the onset of 
flare reconnection. We explore diagnostics of the CME evolution in
the early phase with analysis of the gradual dimming signatures prior
to the CME eruption.
\end{abstract}

\keywords{magnetic reconnection --- Sun: flares --- Sun: coronal mass ejection --- Sun: UV radiation}

\section{Introduction}
\label{intro}
Coronal dimmings were first observed by {\em Skylab} \citep{rust76,rust83}. They 
are usually associated with solar eruptions like coronal mass ejections (CMEs) and flares 
\citep{hudson96,sterling97,harrison03,zhukov04}. Most coronal dimmings are caused by density decrease
due to the loss or rapid expansion of the overlying corona such as by a CME \citep{hudson98,harrison00,zhukov04}. 
This is supported by imaging observations of simultaneous and co-spatial dimming in several coronal lines \citep[e.g.,][]{zarro99,sterling00}, as well as
spectroscopic observations \citep{harra01,tian12}. Therefore, coronal dimming signatures have been analyzed
to infer CME properties including the mass and velocity \citep{Aschwanden2016}.

The more standard approach to study CME evolution and structure has been through analysis 
of coronagraph observations of CMEs, particularly those achieved from multiple view points 
\citep{subramanian14,thernisien09,gopalswamy12,
poomvises10,kilpua12,colaninno13,Cheng2016}. \citet{Aschwanden2016} has recently measured CME properties
by analyzing a few hundred coronal dimming events observed by the {\em Atmospheric Imaging Assembly} \citep[{\em AIA};][]{Lemen2012},
in comparison with the coronagraph observations of CMEs. 

In most of these studies, the CME is detected in the wake of its eruption, and significant coronal
dimming occurs after the flare \citep{Aschwanden2016}. There is much interest in unraveling the
early stage of the CME prior to its explosive release \citep{gopalswamy2006}. \citet{gopalswamy99}
has observed a two-stage coronal dimming as a signature of the CME, which clearly precedes the accompanying
flare. In this study, we report and analyze a two-stage dimming in an eruptive event on 2011 December 26, and
infer the CME expansion in the early phase. \citet{Cheng2016} have given a panoramic view of the flare, CME,
and dimming in this event. They have found that significant dimming occurs at the onset of the flare, 
when CME starts to rise quickly. They have noted that, in this event, gradual dimming is also observed before the onset of the flare 
and the fast rise of the CME, and these pre-flare dimming signatures may help diagnose the CME motion before its fast rise.

In this paper, we focus on these gradual dimming signatures observed from about one hour before the flare onset.
Different from \citet{Cheng2016} that primarily analyze dimming signatures in the
EUV 193\AA\ and 171\AA\ passbands, in this study, we track coronal dimming in the EUV 304\AA\ passband,
which is dominated by emission in the upper chromosphere or transition region. It has been reported that
dimming in coronal EUV lines appears to encompass more extended area than observed in chromsphere lines (e.g., Harvey and Recely 2002, 
and private communication with Terry Forbes). This may be caused by the projection effect of coronal loops 
observed in the coronal lines: when coronal loops are removed or re-oriented away from the line of sight, 
it produces apparent dimming \citep{Qiu2007, Harra2007, Hock2012, Downs2015}. For 
this reason, tracking dimming using the chromosphere lines may minimize the effect of coronal loop projection.
Furthermore, when magnetic reconnection takes place, the chromosphere emission will be enhanced promptly
at the feet of reconnecting field lines. Therefore, observations in the EUV 304\AA\ passband can also reveal
signatures of magnetic reconnection, which is often related to the CME kinematics during its evolution.
In the following text, we will present observations of gradual dimming and transient brightening
in the EUV 304\AA\ passband (Section 2), and infer expansion of the CME magnetic structure in the early phase (Section 3).
Conclusions are given in the last section.

\section{Evolution of Gradual Dimming}

\begin{figure*}[htb]
\begin{center}
\includegraphics[width=16cm]{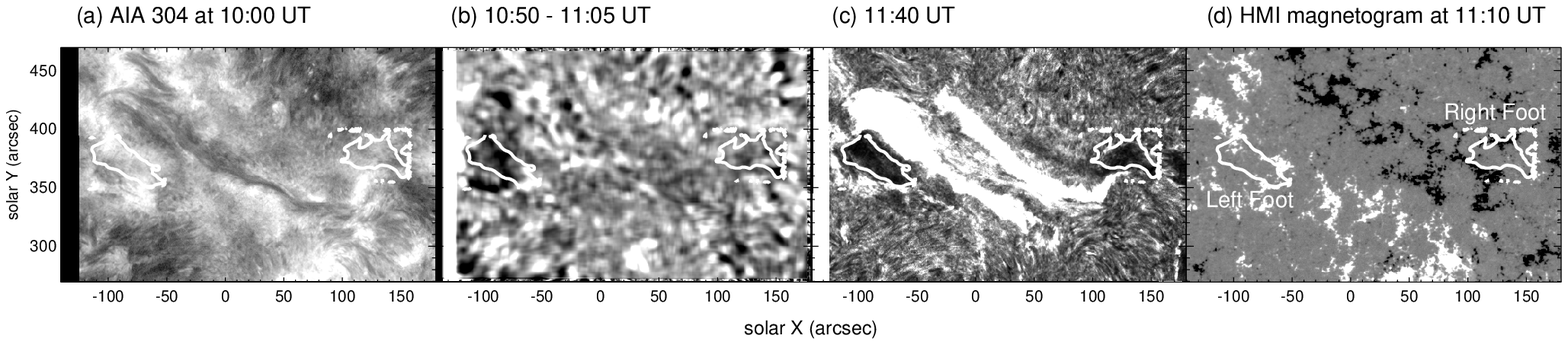}
\caption{Evolution of coronal dimming. (a) {\em AIA} observation of the active region in EUV 304~\AA\ passband
at 10:00~UT. (b) EUV 304~\AA\ image averaged from 10:50 to 11:05~UT and normalized to the image in (a), showing
gradual dimming. (c) EUV 304~\AA\ image at 11:40 UT normalized to the image in (a), showing rapid
dimming and flare ribbons. (d) A longitudinal magnetogram taken by {\em HMI}. In all panels, the contours
indicate the two dimming regions with the brightness of 0.8 (normalized to the brightness at 10:00 UT)
at the times of the maximum dimming area in the Left Foot (LF) and Right Foot (RF), respectively (see text).}
\label{fig1}
\end{center}
\end{figure*}

\begin{figure*}[htb]
\begin{center}
\includegraphics[width=16cm]{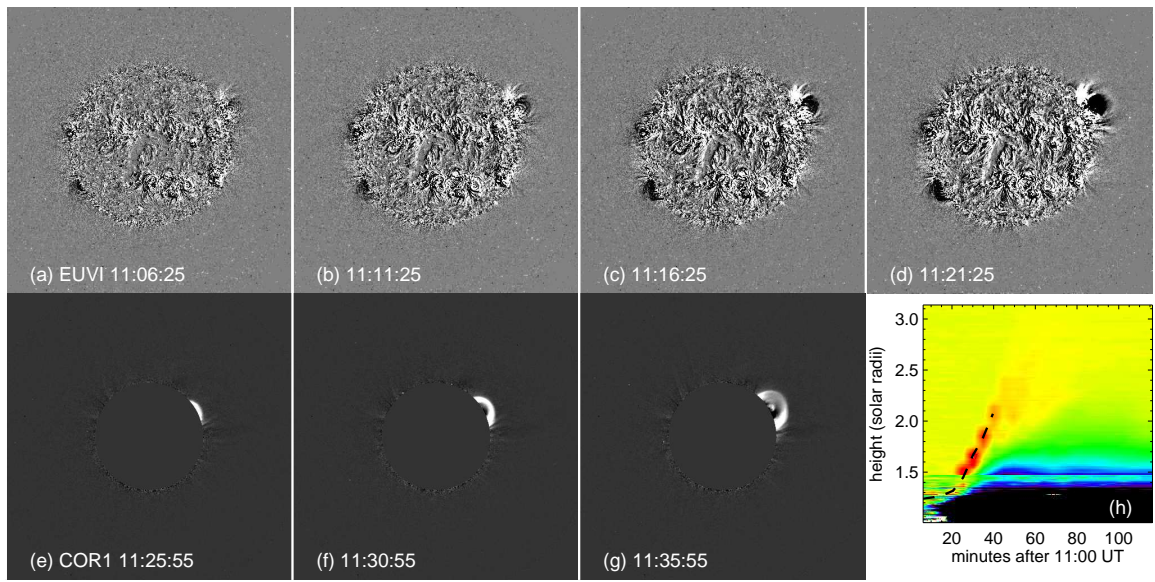}
\caption{Base difference images from observations by EUVI (a-d) and COR1 (e-g) of {\em STEREO-B} showing
the CME from 11:06 to 11:35~UT. (h) Time-height diagram constructed along a slit crossing the solar disk
center and the CME front. The dashed curve indicates the height of the CME front measured by \citet{Cheng2016}.}
\label{cme}
\end{center}
\end{figure*}

An eruptive C5.7 two-ribbon flare occurred on 2011 December 26, and was accompanied 
by a fast Coronal Mass Ejection (CME) and significant coronal dimming \citep{Cheng2016}.
The flare and coronal dimming took place in the active region NOAA 11384 (N13W14) near the disk center, and
were observed by {\em AIA}, as shown in Figure~\ref{fig1}. The CME was observed by the 
{\em Solar Terrestrial Relations Observatory} ({\em STEREO}), as shown in Figure~\ref{cme}.
These observations show that significant dimming occurs at the onset of the impulsive enhancement 
of flare emission as well as the fast rise of the CME at 11:10~UT. The event also
exhibits gradual dimming starting more than half an hour before the flare onset.

Coronal dimming in this event is observed in the EUV 304~\AA\ passband. Figure~\ref{fig1} gives
three snapshots taken in this passband, showing the active region at 10:00~UT (70 min before 
the flare onset at 11:10~UT), the weak gradual dimming before the flare onset, and the 
rapid deep dimming after the flare onset, respectively. The two images in panels b and c 
are each normalized to the image at 10:00~UT (Figure~\ref{fig1}a). 
Two regions at the opposite ends of the two flare ribbons exhibit significant dimming after 
the flare onset (Figure~\ref{fig1}c), 
and weak dimming is also seen in these regions before the flare onset (Figure~\ref{fig1}b). These two regions are referred 
as the Left Foot (LF) and Right Foot (RF), respectively. Overplotted on the images are contours 
at 0.8 (normalized to the brightness at 10:00~UT) at the time when dimming attains the maximum area
in the LF and RF. These contours are used to define the area of the dimming.

Figure~\ref{fig1}d shows the two dimming regions (the 0.8 contours) superimposed 
on a longitudinal magnetogram obtained from the {\em Helioseismic and Magnetic Imager}
\citep[HMI; ][]{Schou2012}.
The contour in the RF covers predominantly negative magnetic flux 
of the amount 8.5$\times 10^{19}$ Mx, and the contour in the LF has positive magnetic flux of
1.1$\times 10^{20}$ Mx. Therefore, the two regions are each monopolar, and
have rather balanced fluxes of opposite polarities. This is consistent with the idea that 
twin dimming maps the conjugate feet of the CME magnetic structure \citep{Webb2000}. 

\begin{figure*}[htb]
\begin{center}
\includegraphics[width=16cm]{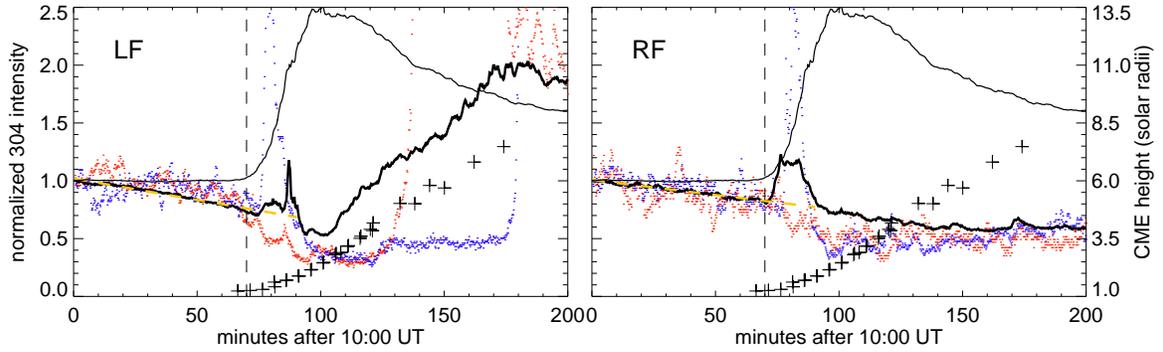}
\caption{Light curves of dimming pixels in the LF (left panel)
and RF (right panel) dimming regions. In each panel, blue and red curves show 
samples of individual pixels, the dark solid curve shows the total light 
curve of all dimming pixels in the region, and the orange guide line shows the fit 
of the dimming light curve to a linear function of time.
The thin black curve shows the total light curve integrated in the entire active 
region and normalized to 10 UT, and the symbols show the CME height
reproduced from the measurements by \citet{Cheng2016}. The vertical dashed lines
indicate the time of flare onset at 11:10 UT. }
\label{lgtcv}
\end{center}
\end{figure*}

Figure~\ref{lgtcv} shows light curves in the 304~\AA\ passband in some sample pixels,
as well as the sum of all the pixels, in each of the two dimming regions.
The total light curve of the active region in the 304\AA\ passband
is also plotted in thin lines. Vertical dashed lines indicate the time of the flare 
onset at 11:10~UT, when the EUV 304\AA\ emission in the active region starts to rise. 
Also plotted in symbols is the height of the CME measured from {\em STEREO} 
observations \citep[][ also see Figure~\ref{cme}h]{Cheng2016}.
The light curves of the dimming regions show that significant dimming occurs 
at or after the flare onset, when the CME visible in the {\em STEREO} field of view 
starts to rise quickly. Most of the pixels also demonstrate prolonged gradual dimming before the flare onset. This gradual dimming
evolution is most clearly illustrated in the total light curve of each of the dimming regions.

Before 11:00 UT, the gradual dimming evolves nearly as a linear function of time, and may be 
approximated by $I/I_0 \sim (\beta - \alpha t)$, where $I$ is the brightness (in units of counts),
$I_0$ is the brightness at 10~UT, $t$ is the time lapse (in units of minutes) from 10:00~UT, and 
the slope $\alpha$ is an indication of the dimming growth rate, in units of percentage 
(with respect to data counts at 10 UT) per minute. We fit the EUV 304 light curve for each pixel 
in the dimming regions to this linear function of time. The majority of the pixels in the 
dimming regions (70\% and 68\% in the LF and RF, respectively) exhibit pre-flare dimming 
with a positive $\alpha$ of up to 0.01 (percentage per minute). We also fit the total light 
curve in each of the two dimming regions, finding the slope $\alpha = 0.004 \pm 0.001, 0.003 \pm 0.001$ 
for the LF and RF, respectively. The other fitting parameter $\beta = 1.007 \pm 0.001, 1.019 \pm 0.001$ 
is very close to unity. In each panel of Figure~\ref{lgtcv}, the orange line 
shows the fit of the total light curve to the linear function of time. 

\begin{figure*}[htb]
\begin{center}
\includegraphics[width=15cm]{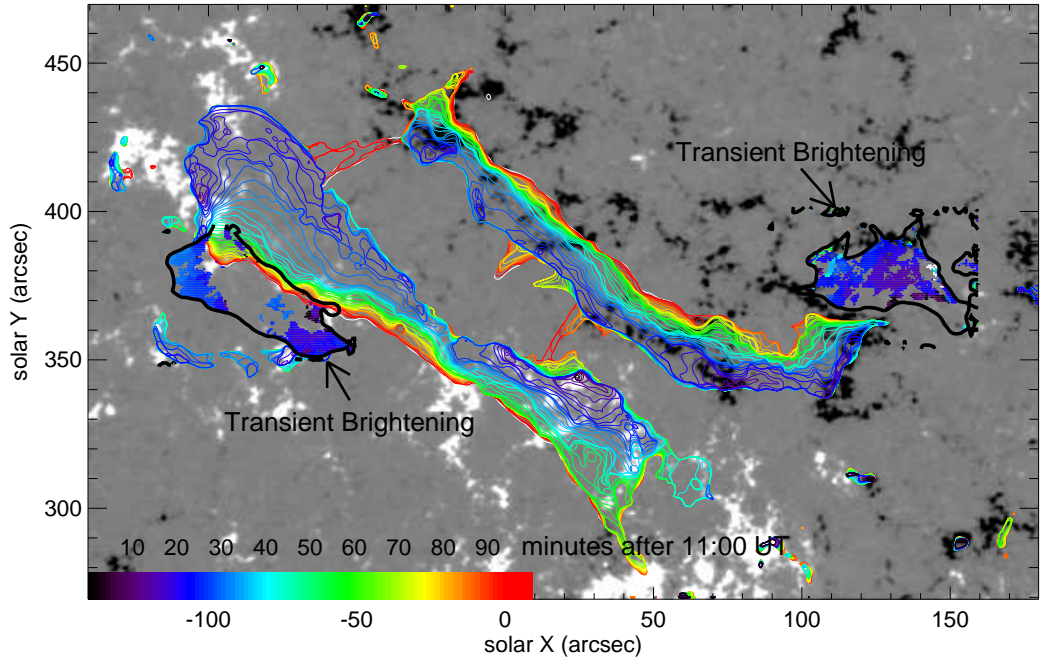}
\caption{Locations of transient brightenings (color symbols) inside dimming regions (marked by thick dark contours),
and flare ribbons (color contours) mapped on a longitudinal magnetogram. The colors of the symbols indicate the time
of the peak transient brightening at a given location, which occurs between 11:10 and 11:30 ~UT. 
The colors of the contours indicate the time when flare ribbons are brightened, showing that 
the inner parts of the ribbons are brightened first at 11:10~UT, and brightening spreads outward for more than one hour.} 
\label{map}
\end{center}
\end{figure*}

Figure~\ref{lgtcv} also reveals transient brightening in the EUV 304\AA\ passband around 
the onset of the flare in some places in the dimming regions. Pixels exhibiting transient 
brightening are mapped in Figure~\ref{map} with color indicating the time when transient 
brightening reaches the peak brightness. About one third to one half of the dimming pixels
exhibit transient brightening, and their brightness peaks within 20 minutes after the flare 
onset at 11:10~UT. The figure also shows the flare ribbon evolution in color contours 
that indicate the time when flare ribbons start to brighten. At the flare onset, 11:10~UT, 
the inner edges of the two ribbons are brightened (violet color), indicating the start of flare 
reconnection forming closed flare loops near the magnetic polarity inversion line. 
Transient brightening occurs around the same time, but in the dimming regions at the outmost
edges of the flare ribbons. Note that whereas the flare ribbon brightening is persistent, 
the transient brightening in the dimming regions is immediately followed by rapid dimming.
In about 20 minutes after the rapid dimming, parts of the LF region are brightened again, due 
to reconnection proceeding away from the polarity inversion line into the dimming region. 
The RF region, on the other hand, is not brightened again and dimming continues for more than an hour. 

\citet{gopalswamy99} have studied an eruptive event, also noting a two-stage
coronal dimming as a signature of the CME. They found weak coronal dimming starting two hours before the rise of the 
flare soft X-ray emission, followed by rapid dimming starting 45 minutes before the flare emission.
The rapid dimming is considered to reflect the onset of the CME, which well precedes the flare
in that event. Similarly, the observed persistent gradual dimming in this event may
indicate the slow rise of the CME magnetic structure starting an hour before
the flare, and the onset of the CME as observed by {\em STEREO} also coincides 
with the rapid dimming. But different from the case reported by \citet{gopalswamy99}, 
in this event, the rapid dimming and the fast rise of the CME take place at or after the flare onset. 
Furthermore, transient brightenings are observed in the dimming regions at the flare onset, 
which mark the transition between the two stages of dimming. It is likely that transient brightenings 
in the dimming regions are produced by reconnection of the CME structure itself with overlying field lines, 
which, together with the tether-cutting reconnection below the CME structure forming flare ribbons, enables 
eruption of the CME, and causes subsequent rapid dimming at the feet.

\section{CME Expansion In the Early Phase}

To infer the rise of the CME in the early phase, we model the observed gradual 
dimming in the EUV 304~\AA\ passband, assuming that emission in this passband by the transition region or upper chromosphere
is proportional to the gas pressure $P$ \citep[the so-called pressure gauge; ]
[]{Fisher1987, Hawley1992, Qiu2013, Zeng2014}, which is uniform from the corona to the transition region 
and upper chromosphere in quasi-static equilibrium. 
Therefore, the time evolution of the 304\AA\ emission at the base of the corona follows the 
simple relation $I(t) \sim P(t)$.


When the CME coronal structure expands, it reduces the plasma density $n$ and/or temperature $T$,
and therefore the gas pressure $P \sim nT$ decreases. To study this change, a few CME expansion
models are adopted to calculate the evolution of the coronal pressure. 
In general, these models assume 1d or self-similar CME expansion, either isothermally or adiabatically
\citep{Aschwanden2016}. Mass conservation further requires $nL$ (1d expansion) or 
$nL^3$ (self-similar expansion) to be constant, where $L$ is the length-scale of the 
CME structure. We assume $L \sim H$, where $H$ is the height of the coronal structure above 
the solar surface anchored at the dimming regions. \footnote{Note that the curvature effect is ignored for the early-phase CME at a low height relative to the Sun's radius.}

With these considerations, evolution of the EUV 304\AA\ emission can be described
by $$ \frac{I}{I_0} = \frac{P}{P_0} = \frac{nT}{n_0T_0} = \left(\frac{H}{H_0}\right)^{-1/\eta}, $$
where $I_0, P_0, n_0$ and $T_0$ refer to properties at the initial time, and
$\eta$ depends on the expansion model. With the ratio of specific heats $\gamma = 5/3$, 
the power index is given by $\eta = 1, 1/3, 3/5, 1/5$ for the 1d isothermal 
expansion (IS1 hereafter), self-similar isothermal expansion (IS3), 1d adiabatic expansion (AD1), 
and self-similar adiabatic expansion (AD3), respectively.
From observations, $I/I_0 = (1 - \alpha t)$, the height and speed of the expanding plasma 
rooted at the dimming region can be computed as $ H = H_0 (1-\alpha t)^{-\eta} , $ and 
$ v = H_0 (\alpha \eta) (1-\alpha t) ^{-\eta -1}$, respectively.

\begin{figure*}[htb]
\begin{center}
\includegraphics[width=15cm]{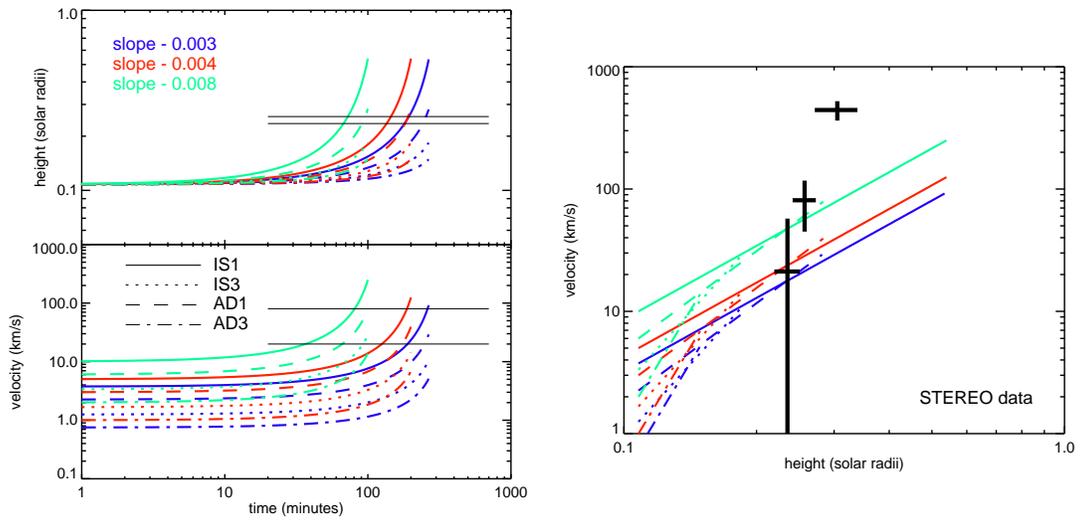}
\caption{Left: height (top) and speed (bottom) calculated with the 4 expansion models using
three dimming rates. The two horizontal bars in the plots indicate the height (speed) of the CME observed
by STEREO at the first two times. Right: height-speed diagram from the models. The dark symbols show the
STEREO measured CME height and speed \citep{Cheng2016}.}
\label{fig5}
\end{center}
\end{figure*}

The left panels in Figure~\ref{fig5} shows the evolution of the CME height and speed calculated
with several values of $\alpha$ and with the four models. Since the separation of the two 
dimming regions is 200~\arcsec (Figure~\ref{fig1}), we take the initial height of the CME ($H_0$) 
to be 100~\arcsec, assuming that it is a semi-circular structure.
The values of $\alpha = 0.003, 0.004$ from fitting the light curves of the two dimming regions are used in the calculation.
In addition, we also calculate the case for $\alpha = 0.008$, with the consideration that the observed
dimming rate may be smaller than the actual dimming rate due to effects of scattered light and uncorrected
point spread function. The height and speed curves have asymptotical behavior as $I/I_0$ approaches zero, when the
relation in the above equations is no longer valid. 
We carry the calculation to the time when $I/I_0 = 0.2$. It is apparent that the CME height, as well as its speed, is largest by
the 1d isothermal expansion, followed by the 1d adiabatic, 3d isothermal, and 3d adiabatic models.
The right panel shows the height-speed diagram of the four models with the three dimming rates;
the growth rate of the speed with respect to the height is largest in AD3 model, followed by IS3, AD1, and IS1
models.

It is interesting to compare these curves with the {\em STEREO} observations of the CME. 
The CME becomes visible in the {\em STEREO-B EUVI} field of view at 11:06 UT (Figure~\ref{cme}), 
when its height is measured to be $0.24\pm0.02$ solar radii above the surface, and its 
speed, which is the time derivative of the height, is $21\pm36$~km s$^{-1}$. 
In the next time at 11:11 UT, its height is $0.26\pm 0.02$ solar radii, 
its speed being $81\pm36$ km~s$^{-1}$. The dark symbols in the height-speed diagram 
in Figure~\ref{fig5} show the measurements from the {\em STEREO} observations
by \citet[][ also reproduced in Figure~\ref{lgtcv}]{Cheng2016}, and the vertical and horizontal bars mark the uncertainties. 
Comparing these measurements with the model calculation, we find the first
measurement of the CME height and speed at 11:06 UT is well within the 
range of the 1d models, and the second measurement at 11:11 UT is close to the 
upper-limit of the 1d models. Note that these two measurements are obtained at 
about 70 minutes after the initial time at 10:00~UT. In the height and speed plots 
in the left panels of Figure~\ref{fig5}, two horizontal lines mark the first two measurements. 
It is seen that, with the observed dimming rate ($\alpha = 0.003, 0.004$) and 
prescribed initial height $H_0$, it takes more than 100 minutes to reach the observed 
CME height following the two 1d expansion models. With a reasonably larger initial height $H_0$ or a 
larger dimming rate $\alpha$, the observed CME height and speed can be
produced in a shorter time. On the other hand, the two 3d models cannot produce results
close to the first two {\em STEREO} measurements even if we vary the initial model parameters in 
a reasonable range.

Five minutes later, at 11:16 UT, the CME is at 0.31 solar radii with a speed of a few hundred kilometers per second.
The diagram in Figure~\ref{fig5} shows that the results from the 1d models are no longer comparable with the CME speed
at this height, suggesting that CME kinematics has changed from the gradual evolution, most likely 
because of the reconnection onset at 11:10 UT. Meanwhile, rapid dimming is observed at this
time reflecting the rapid expansion of the CME.

\section{Conclusion \& Discussions}
\label{discussion}

In this paper, we have shown that the twin dimming regions in an eruptive event reported in \citet{Cheng2016} also exhibit
gradual dimming in the EUV 304\AA\ passband from more than half an hour before the flare onset
and fast rise of the CME. Transient brightening in this passband also occurs in the dimming regions
at the time when flare ribbons start to form. The transient brightening is followed by 
rapid dimming coincident with the fast rise of the CME. It is likely that CME field lines reconnect
with the ambient field, which produces the transient brightening at the feet \citep{Downs2015},
and flare ribbons are produced by reconnection below the CME. Therefore, reconnection marks 
the transition between the two stages of dimming, or the two stages of the CME evolution. 

We analyze evolution of the gradual dimming, which grows nearly linearly with time,
to infer the CME height and speed in the early phase, with the assumption that the 304 counts rate 
is proportional to the coronal pressure. We find that, with the observed dimming rate, 
the radial (1d) expansion model can best match the measurements of the CME height and speed 
when the CME is first observed by {\em STEREO} from the limb. Shortly after the onset 
of reconnection, the kinematic evolution of the CME has significantly changed from the gradual evolution. 

The analysis in this study shows the potential that high-cadence
disk observations with high-quality photometry by {\em AIA} may be used
to reconstruct evolution of the CME (if pre-existing) in the early phase
before its catastrophic eruption. 


\acknowledgments The authors thank the referee for critical comments that help improve the manuscript.
The work is supported by the NSF grant 1460059.  J.C. is  supported by the NSFC under grants 11373023,
11133004, 11103008 and 11673048. \textit{SDO} is a mission of
NASA's Living With a Star Program.

\bibliographystyle{apj}

\end{document}